\documentstyle[aaspp4]{article}

\begin{document}
\newcommand {\ppmm}{P${}^2$M${}^2$}
\newcommand{\Vec}[1]{{\mbox{\boldmath{${#1}$}}}}

\lefthead{Kawai and Makino}
\righthead{A Simple Formulation of FMM: P${}^2$M${}^2$}
\title{A Simple Formulation of the Fast Multipole Method:
Pseudo-Particle Multipole Method}
\author{Atsushi Kawai$^{1}$
\and
Junichiro Makino$^{1}$
}

\affil{
$^1${
	Department of General Systems Studies,
	College of Arts and Sciences,
	University of Tokyo,
}\\
{
	3-8-1 Komaba,
	Meguro-ku,
	Tokyo 153
}\\
}
\date{}
\authoremail{kawai@grape.c.u-tokyo.ac.jp}

\begin{abstract}
We present the pseudo-particle multipole method (\ppmm), a new method
to handle multipole expansion in fast multipole method and
treecode. This method uses a small number of pseudo-particles to
express multipole expansion. With this method, the implementation of
FMM and treecode with high-order multipole terms is greatly
simplified. We applied \ppmm ~to treecode and combined it with
special-purpose computer GRAPE. Extensive tests on the accuracy and
calculation cost demonstrate that the new method is quite attractive.
\end{abstract}

\section{Introduction}

In this paper, we describe the pseudo-particle multipole method
(\ppmm, \cite{m98}), a new method to express multipole expansion in
fast multipole method(FMM, \cite{gr87}) and treecode(\cite{bh86}).
The basic idea of \ppmm ~is to use a small number of pseudo-particles
to express the multipole expansion. In other words, this method
approximates the potential field of physical particles by the field
generated by a small number of pseudo-particles. The distribution of
pseudo-particles is determined so that it correctly describes the
coefficients of multipole expansion.

\ppmm ~offers two advantages over standard FMM which uses the
multipole expansion directly. One advantage is its simplicity. As will
be seen, the translation formulae used in \ppmm ~are much simpler
than their counterpart in standard FMM.  Although the calculation
cost is roughly the same for the same level of the accuracy, the
simplicity implies that performance tuning and parallelization are
easier.

Another advantage is that \ppmm can take full advantage of
special-purpose computers(GRAPE, \cite{mt98}; \cite{s90}). GRAPE is a
pipelined processor specialized to direct force calculation between
particles. It offers the price-performance 100-1000 times better than
that of general-purpose computers, for direct force
calculation. Though treecode has been used on
GRAPE(\cite{at98}; \cite{f91}; \cite{m91}), it
was difficult to go to high accuracy since only monopole can be
calculated on GRAPE. With \ppmm ~we can evaluate high-order terms
using GRAPE, since these terms are expressed by distribution of
pseudo-particles.

This paper is organized as follows.
In section \ref{sec:treefmm} we briefly describe the treecode and the
FMM. In section \ref{sec:am} we give the description of Anderson's
method (\cite{a92}), to which our method is closely related.  In section
\ref{sec:ppmm} we describe the mathematics of \ppmm.  In section
\ref{sec:accppmm} we present the result of numerical tests on the
accuracy and the calculation cost for our implementation of treecode
with \ppmm. In section \ref{sec:grape} we discuss the implementation
of treecode with \ppmm ~on GRAPE. In section \ref{sec:ppmmam} we
compare \ppmm ~with Anderson's method. In section \ref{sec:summary} we
summarize this paper.

\section{Treecode and FMM}\label{sec:treefmm}

Here we give brief description of treecode and FMM. The treecode is
described in section \ref{sec:tree} and the FMM is described in
section \ref{sec:fmm}.

\subsection{Treecode}\label{sec:tree}

In treecode, the forces from a group of distant particles are
approximated by multipole expansions. Hierarchical tree structure is
used for grouping of the particles.

Here we summarize the calculation procedure following \cite{h87};
\cite{hk89}.
%
%
First we construct an oct-tree structure by hierarchical subdivision
of the space. The division procedure starts from the root node(root
cell) of the tree which corresponds to a cube covering the entire
system. The procedure is repeated until all leaf cells contain only
one or zero particles.

In the next step we calculate the multipole expansions for all
non-leaf cells. The calculation begins from the parents of the leaf
cells and continued to the root cells ascending the tree structure.
For a cell whose children are all leaves, the multipole expansion at
its center is directly calculated from the distribution of the
particles(leaves) in it. The expansion of a higher level cell is
calculated from the expansions of its children. For each child cell,
the center of the expansion is shifted to the center of the parent(M2M
shift). All shifted expansions are then summed up at the center of the
parent cell. Note that in most of existing implementation of treecode
only up to quadrupole term is retained, and the center of mass is used
as the center of expansion.

Then we calculate the total force on each particle. Starting from the
root cell, we recursively traverse the tree structure collecting the
force from cells(\cite{bh86}). We examine whether the cell in question
is well separated from the particle or not. If the cell is well
separated, the multipole expansion of the cell is evaluated at the
position of the particle and added to the total force on the particle. 
In the case of a leaf cell, the force from the particle in it is used
instead of the multipole expansion. If the cell is not well separated,
we descend the tree to resolve the current cell into child cells, an
then recursively examine each child in the same way. The condition
that a cell is well separated is expressed as $l/d < \theta$. Here $l$
is the side length of the cell, $d$ is the distance between the cell
and the particle, and $\theta$ is the opening angle that controls the
accuracy. Leaf cells are always considered to be well separated.

The calculation cost of treecode is $O(N \log N)$, since the particle
sees larger cell as distance becomes larger.

Treecode is widely used in astrophysics, in particular where accuracy
requirement is modest. Most of existing implementations of treecode
use only up to quadrupole moment and calculation cost rises quickly
when high accuracy is required.
The implementation detail is given in \cite{h87}; \cite{hk89}. For
the implementation on distributed-memory parallel computers, see
\cite{s94}.

\subsection{FMM}\label{sec:fmm}

\begin{figure}[htbp]
\plotone{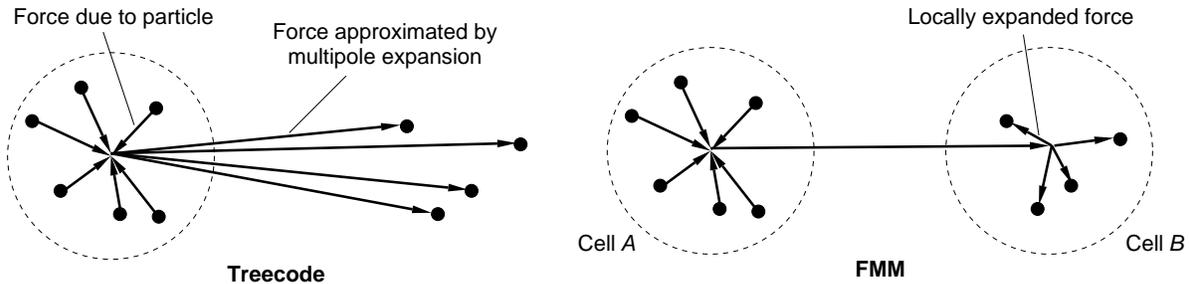}
\caption{Approximation used in treecode(left) and FMM(right).
}\label{fig:treefmm}
\end{figure}

Figure \ref{fig:treefmm} shows the conceptual difference between
treecode and FMM. In treecode, the forces from a cell to different
particles are evaluated independently. In FMM, on the other hand, to
calculate forces from cell $A$ to particles in cell $B$, we first
calculate the spherical harmonics expansion(local expansion) of the
potential field at the center of cell $B$ and then evaluate that
expansion at the position of particles in cell $B$.

FMM was first presented in \cite{gr87} for two dimensional case and
was extended to three dimension in \cite{gr88}.  The implementation
detail for the three dimensional case is given in \cite{b94};
\cite{gr88}; \cite{hj96}; \cite{pg96}; \cite{sl91}.

%

In the following we describe the non-adaptive version of FMM for three
dimensional case.
First we construct an oct-tree structure by hierarchical subdivision
of the space. The division procedure starts from the root cell at
refinement level $R=0$, which covers the entire system.  Here we
define the refinement level $R$ as the depth of the tree. We repeat
the procedure until a given refinement level $R_{\rm max}$ is reached.
The level $R_{\rm max}$ is chosen so that the average number of
particles in one leaf cell roughly equals the prescribed number which
is determined to optimize the calculation speed.

In the next step, we calculate the multipole expansions for all cells. 
This part is the same as that for treecode described in section
\ref{sec:tree}.

\begin{figure}[htbp]
\plotone{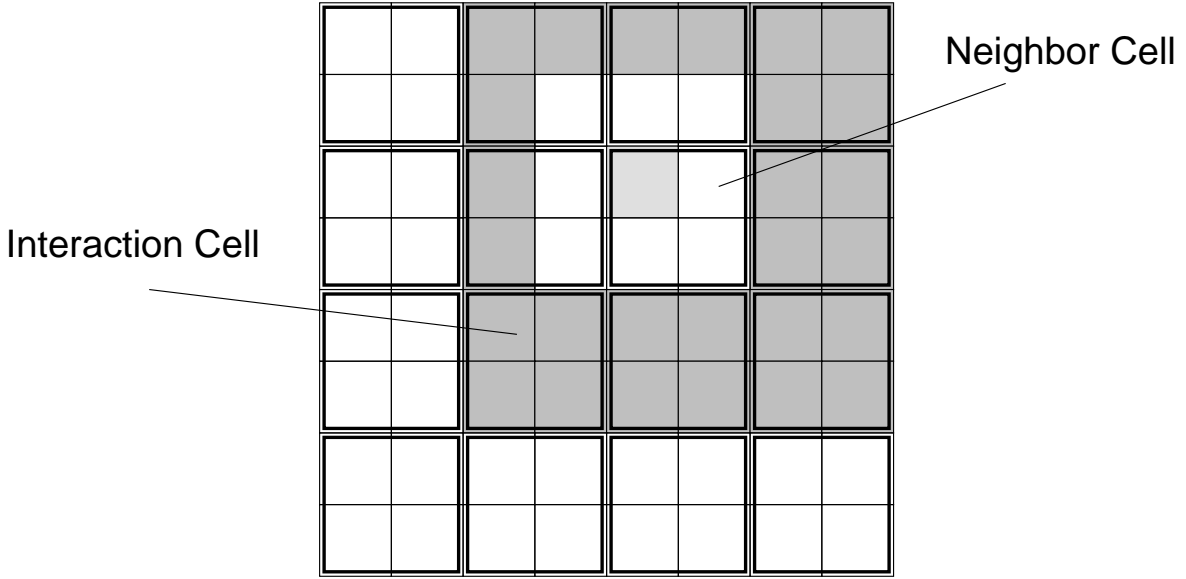}
\caption{Relation between interaction cells(shaded) and neighbor cells
of a hatched cell.
}\label{fig:fmmrange}
\end{figure}

Then, for each cell, we calculate the local expansion due to its
interaction cells. Figure \ref{fig:fmmrange} shows the relation
between a cell and its interaction cells. Interaction cells of a cell
are defined as the children of its parent's neighbor cells which are
not its own neighbors. Here, neighbor cells are the cells at the same
level which are in contact with the cell.
The contribution from the interaction cells are calculated by
converting their multipole expansion to the local expansions at the
center of the objective cell(M2L conversion), and then summing them
up.

In the next step, we add up the local expansions at different levels
to obtain the total potential field at the leaf cells. We start
calculation at $R=3$. For all cells in $R=3$, we shift the center of
the local expansion of its parent(L2L shift), and then add it to the
local expansion of the cell. By this procedure, all cells in $R=3$
will have the local expansion of the total potential field except for
the contribution of the neighbor cells. By repeating this procedure
for all levels, we obtain the potential field for all leaf cells.

%

Finally, we calculate the total force on each particle. The total
force is calculated as a sum of the distant and the neighbor
contributions. The distant part is calculated by evaluating the local
expansion of the leaf cell at the position of the particle. The
neighbor part is directly calculated by evaluating the
particle-particle forces.

In the following, we summarize the mathematics used in FMM.
FMM requires two types of expansion of the potential and three
transformations of them.  One of the two expansions is the multipole
expansion and the other is the local expansion.
The multipole expansion of the potential $\Phi(\Vec{r})$ up to $p$-th
order is expressed as
\begin{equation}\label{eq:mpe}
  \Phi(\Vec{r}) = \sum_{l=0}^p \sum_{m=-l}^{m=l}
                  \frac{\alpha_l^m}{r^{l+1}}Y_l^m(\theta, \phi)
\end{equation}
in spherical coordinates $\Vec{r}(r, \theta, \phi)$.  Here
$Y_l^m(\theta,\phi)$ is the spherical harmonics and $\alpha_l^m$ are the
expansion coefficients. In order to approximate the potential field
due to the distribution of particles, the coefficients should satisfy
\begin{equation}\label{eq:coeff}
  \alpha_l^m = \frac{4\pi}{2l+1} \sum_{i=1}^N m_i r_i^l
               Y_l^{m*}(\theta_i, \phi_i),
\end{equation}
where $N$ is the number of particles to be approximated, $m_i$ and
$(r_i, \theta_i, \phi_i)$ are the masses and positions of the
particles, and * denotes the complex conjugate.
The local expansion $\Psi(\Vec{r})$ up to $p$-th order is given by
\begin{equation}\label{eq:le}
  \Psi(\Vec{r}) = 4 \pi \sum_{l=0}^p \sum_{m=-l}^{m=l}
                  \beta_l^m r^lY_l^m(\theta, \phi),
\end{equation}
where $\beta_l^m$ is the expansion coefficients.

The three transformation required for FMM are the M2M shift, the M2L
conversion, and the L2L shift. These transformation are expressed as
follows:
\begin{eqnarray}
  \label{eq:m2m}
  \alpha_{l'}^{m'} & = & \sum_{l=0}^p \sum_{m=-l}^{m=l}
                     T^{MM}_{l'm',lm} \alpha_{l}^{m}
  \hspace*{30pt}{\rm : M2M},\\
  \label{eq:m2l}
  \beta_{l'}^{m'} & = & \sum_{l=0}^p \sum_{m=-l}^{m=l}
                     T^{LM}_{l'm',lm} \alpha_{l}^{m}
  \hspace{30pt}{\rm : M2L},\\
  \label{eq:l2l}
  \beta_{l'}^{m'} & = & \sum_{l=0}^p \sum_{m=-l}^{m=l}
                     T^{LL}_{l'm',lm} \beta_{l}^{m}
  \hspace{30pt}{\rm : L2L}.
\end{eqnarray}
Using
\begin{equation}
  a_{lm} = (-1)^{l+m} \frac{(2l+1)^{1/2}}{[4 \pi (l+m)!(l-m)!]^{1/2}},
\end{equation}
the transformation matrices are expressed as
\begin{eqnarray}
  \label{eq:tmm}
  T^{MM}_{l'm',lm} & = &
  4 \pi \frac{(-r_{t})^{l'-1}Y^{(m'-m)*}_{l'-l}(\theta_t,\phi_t)
              a_{l'-l,m'-m} a_{lm} (2l'+1)}
             {(2l+1)[2(l'-l)+1]a_{l'm'}}, \\
  \label{eq:tml}
  T^{LM}_{l'm',lm} & = &
  4 \pi \frac{(-1)^{l+m}Y^{(m'-m)*}_{l'+l}(\theta_t,\phi_t)
               a_{lm} a_{l'm'}}
             {r_t^{l'+l+1}(2l+1)(2l'+1)a_{l'+l,m'-m}}, \\
  \label{eq:tll}
  T^{LL}_{l'm',lm} & = &
  4 \pi \frac{r_{t}^{l-l'}Y^{(m-m')*}_{l-l'}(\theta_t,\phi_t)
              a_{l'm'} a_{l-l',m-m'}}
             {a_{lm} (2l'+1)[2(l-l')+1]}.
\end{eqnarray}
Here, $(r_t, \theta_t, \phi_t)$ is the position of the new origin
relative to the old one.

The calculation cost of FMM is $O(N)$. Therefore the scaling of FMM is
better than that of treecode. However, at least in three dimension,
comparisons indicate that treecode is faster for realistic number of
particles (\cite{bs97}).

\section{Anderson's Method}\label{sec:am}

Anderson (\cite{a92}) proposed a simple formulation of FMM based on
Poisson's formula. Poisson's formula gives the solution of the
boundary value problem of the Laplace equation. When potential on the
surface of a sphere of radius $a$ is given, the potential $\Phi$ at
position $\Vec{r}$ is expressed as
\begin{equation}\label{eq:amo}
  \Phi(\Vec{r}) = \frac{1}{4\pi}\int_S\sum_{n=0}^\infty(2n+1)
  \left(\frac{a}{r}\right)^{n+1}
  P_n\left(\frac{\Vec{s}\cdot\Vec{r}}{r}\right)\Phi(a\Vec{s})ds
\end{equation}
for $r > a$, and
\begin{equation}\label{eq:ami}
  \Phi(\Vec{r}) = \frac{1}{4\pi}\int_S\sum_{n=0}^\infty(2n+1)
  \left(\frac{r}{a}\right)^{n}
  P_n\left(\frac{\Vec{s}\cdot\Vec{r}}{r}\right)\Phi(a\Vec{s})ds
\end{equation}
for $r < a$.
Here $\Phi(a\Vec{s})$ is the given potential on the sphere
surface. The range $S$ of the integration covers the surface of a unit
sphere centered at the origin. The function $P_n$ denotes the $n$-th
Legendre polynomial.

In Anderson's method the value of $\Phi(a\Vec{s})$ is used to
express the multipole/local expansion, while the original FMM uses the
coefficients of the expansion terms.
The advantage of Anderson's method over the original FMM is its
simplicity.  The translation operators described in section
\ref{sec:fmm} are all realized by evaluating the potential on points
of the target sphere due to the source sphere.  All such evaluations
are performed using equation(\ref{eq:amo}) and (\ref{eq:ami}). These
are by far simpler to implement the original FMM using
equation(\ref{eq:m2m})-(\ref{eq:tll}).

In order to evaluate the integral in equation(\ref{eq:amo}) and
(\ref{eq:ami}) numerically, we need to sample the values of
$\Phi(a\Vec{s})$ on the surface of the sphere. The number of sample
points required and the optimal positions of such points are given in
\cite{hs96}.


\section{Pseudo-Particle Multipole Method}\label{sec:ppmm}

\ppmm ~is quite similar to Anderson's method. The difference is that
in \ppmm ~we use the mass distribution on the surface of a sphere
instead of the potential. The continuous mass distribution is again
approximated by finite number of pseudo-particles, and the potential
exerted by these pseudo-particles are calculated in the same way as
that exerted by physical particles.

Conceptually, physical particles are converted to pseudo-particles in
the following two steps.  First we calculate multipole expansion. Then
we assign mass to pseudo-particles so that they have the same
multipole expansion as physical particles.

Calculation procedure of these two steps are as follows:
In the first step we expand the potential with spherical harmonics. As
we have seen in section \ref{sec:fmm}, the multipole expansion and its
coefficients are given by equation(\ref{eq:mpe}) and (\ref{eq:coeff}).
In the second step we find a continuous mass distribution
$\rho(a,\theta, \phi)$ on a sphere of radius $a$, which approximate
the potential field. Then we approximate that distribution by
pseudo-particles. The mass distribution should satisfy
\begin{equation}
  \alpha_l^m = \frac{4\pi a^{l+2}}{2l+1}
               \int_S \rho(a, \theta, \phi) Y_l^{m*}(\theta, \phi)ds,
\end{equation}
for $0 \le l \le p$. Here $p$ is the highest order of multipole
expansion to express and $S$ denotes the surface of the unit
sphere. Because the spherical harmonics comprise an orthonormal
system, $\rho(a, \theta, \phi)$ is expressed as
\begin{equation}\label{eq:contmass}
  \rho(a, \theta, \phi) = \sum_{l=0}^p
                          \sum_{m=-l}^{l}
                          \frac{2l+1}{4\pi a^{l+2}}
                          \alpha_l^m Y_l^{m}(\theta,\phi).
\end{equation}
Following \cite{hs96}, we place $K$ points on the sphere so that they
satisfy $\sum^{K}_{j=1}f(\Vec{r_j})/K = \int_S f(a\Vec{s})ds$,
where $f$ is any polynomial of degree at most $[2p+1]$(order
$[2p+1]$ is necessary to guarantee orthogonality of spherical harmonics
of up to order $p$), and $\Vec{r}_j$ are the positions of the
distributed points. Thus equation(\ref{eq:contmass}) is replaced by
\begin{equation}\label{eq:ppm}
  m_j = \frac{4\pi}{K}\sum_{l=0}^p \sum_{m=-l}^{l}
        \frac{2l+1}{4\pi a^l}
        \alpha_l^m Y_l^{m}(\theta_j,\phi_j).
\end{equation}
where $m_j$ are the masses of pseudo-particles located at $\Vec{r}_j
(a,\theta_j, \phi_j)$.
In practice, the masses $m_j$ of pseudo-particles are calculated
directly from the positions $\Vec{r}_i (r_i,\theta_i,\phi_i)$ and
masses $m_i$ of physical particles. Combining equation(\ref{eq:coeff})
and (\ref{eq:ppm}), $m_j$ is expressed as
\begin{equation}\label{eq:ppm2}
  m_j = \frac{4\pi}{K}\sum_{l=0}^p
        \sum_{m=-l}^{l}
        \sum_{i=1}^N m_i
        \left(\frac{r_i}{a}\right)^l
        Y_l^{m}(\theta_j, \phi_j) Y_l^{m*}(\theta_i, \phi_i).
\end{equation}
Applying the addition theorem of spherical harmonics to
equation(\ref{eq:ppm2}), we obtain:
\begin{equation}\label{eq:ppmm}
  m_j = \sum^{N}_{i = 1} m_i
        \sum^{p}_{l=0}\frac{2l+1}{K}
        \left(\frac{r_i}{a}\right)^l
        P_l(\cos \gamma_{ij}),
\end{equation}
where $\gamma_{ij}$ is the angle between $\Vec{r}_i$ and $\Vec{r}_j$.

\section{Accuracy of the Force Calculated with \ppmm}\label{sec:accppmm}

Here we present the result of numerical tests for our implementation
of treecode with \ppmm. In section \ref{sec:acc} we compare the
accuracy of \ppmm ~and the original FMM for single particle. In
section \ref{sec:acc-cost} we present the relation between the
accuracy and the calculation cost for our implementation of treecode
with \ppmm.

\subsection{Accuracy of the Force Exerted from One Particle}\label{sec:acc}

\begin{figure}
\plotone{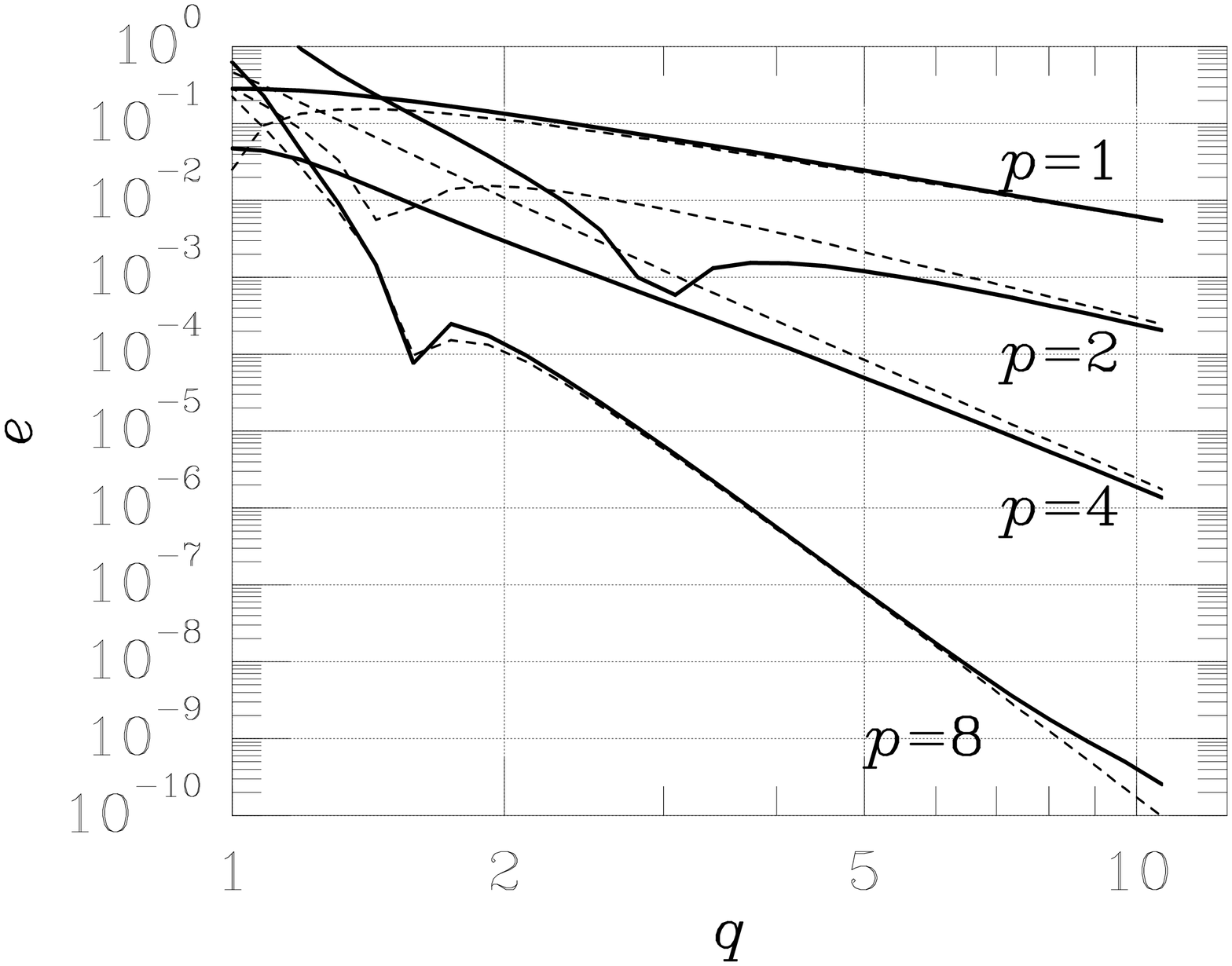}
\caption{The error of potential calculated with \ppmm (solid) and
original multipole expansion (dashed). The error of potential
generated by one particle at {\bf \it p} are plotted against the
distance from the origin to the evaluation point {\bf \it q}. The
error $e$ is defined as the difference from the exact potential. From
top to bottom, four curves are for $p=1$, $2$, $4$ and $8$,
respectively. The sphere radius $a$ is 1.0.
}\label{fig:onecell}
\end{figure}

We measured the accuracy of the potential calculated with \ppmm ~and
the original multipole expansion.  A point mass located at $\Vec{p}
(1,0,0)$ in spherical coordinate is used as the source of the
potential. In figure \ref{fig:onecell} the error of the potential
evaluated at point $\Vec{q} (q,2\pi/3,0)$ is plotted as a function of
$q$ for both \ppmm ~and the original multipole expansion.  We can see
that their agreement is quite good.

\subsection{Accuracy of the Total Force}\label{sec:acc-cost}

\begin{figure}
\plotone{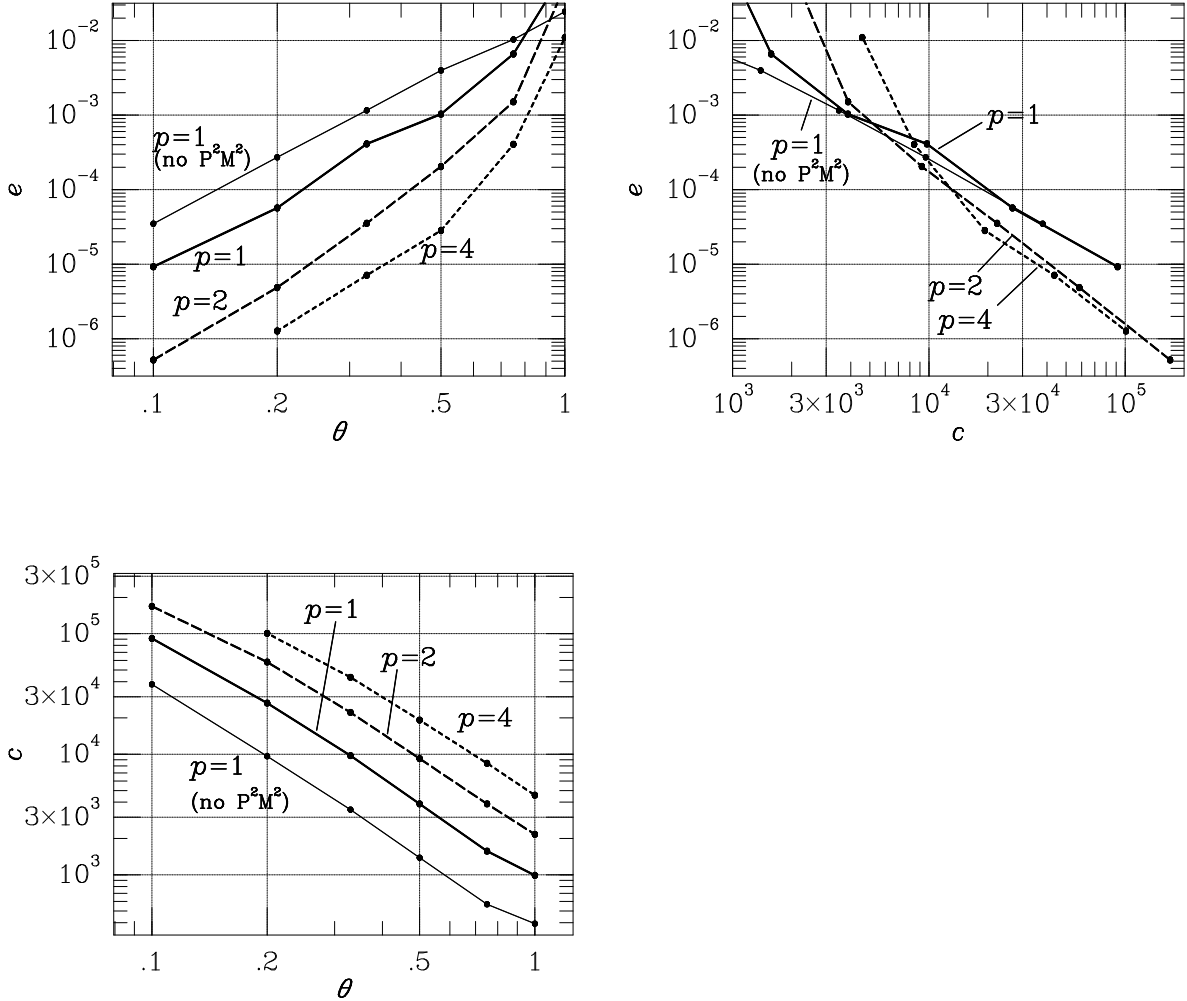}
\caption{The error $e$ plotted against the calculation cost
$c$(top-left panel) and the opening angle $\theta$(top-right
panel). The relation between $c$ and $\theta$ is also shown in the
bottom-left panel. Thick and thin curves denote treecode with and
without \ppmm, respectively. Solid, long dashed, and short dashed
curves are for $p=1$, $2$, and $4$, respectively.
}\label{fig:total}
\end{figure}

We measured the accuracy of the force calculated by the treecode
with \ppmm.
We used 262144 equal-mass particles uniformly distributed in a sphere. 
We measured the relative error $e$ of the force averaged over all the
particles, and the calculation cost $c$ for $p$ = 1, 2, and 4.
The average error $e$ is defined as
\begin{equation}
  e^2 =
  \frac{1}{N}\sum^N_{i=1}\frac{|\hat\Vec{f}_i-\Vec{f}_i|}{|\Vec{f}_i|}^2,
\end{equation}
where $\hat\Vec{f}_i$ is the force calculated with treecode and
$\Vec{f}_i$ is the force calculated with direct summation method. We
define the cost $c$ as the average number of (pseudo)-particle-particle
interactions evaluated for one particle.
Figure \ref{fig:total} shows the results. We can see that \ppmm
~version with high-order terms is faster than the original version with
monopole term, when high accuracy ($e < 10^{-4}$) is required.

\section{\ppmm ~Treecode on GRAPE}\label{sec:grape}

\begin{figure}
\plotone{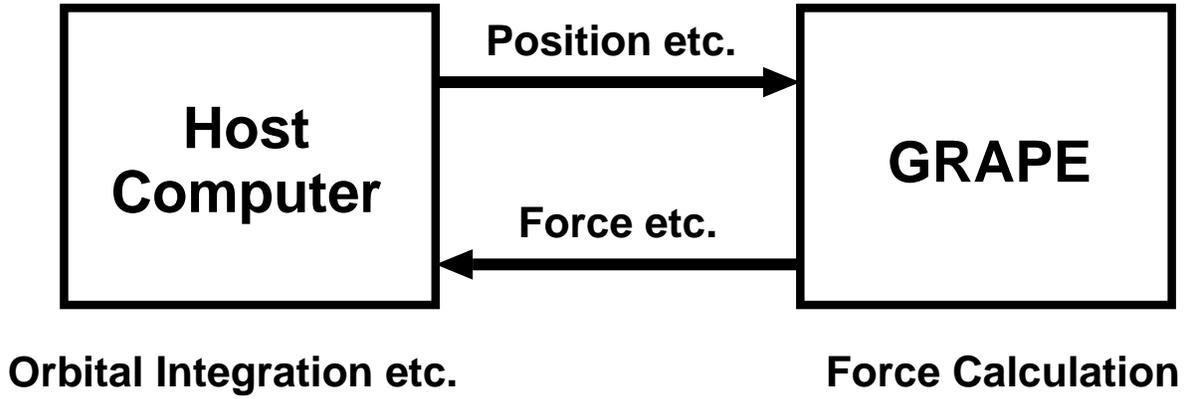}
\caption{The GRAPE system.}\label{fig:grape}
\end{figure}

We implemented the treecode with \ppmm ~on GRAPE (\cite{mt98};
\cite{s90}).
GRAPE is a special-purpose computer for gravitational $N$-body
problem(figure \ref{fig:grape}). It consists of general-purpose host
and special-purpose backend which calculates particle-particle
interaction.

With \ppmm, high-order terms of multipole expansion are expressed as
particles. Therefore we can implement FMM and treecode with arbitrary
order of multipole expansion on GRAPE systems. Previously, we could
use GRAPE only to evaluate monopole terms.

\begin{figure}
\plottwo{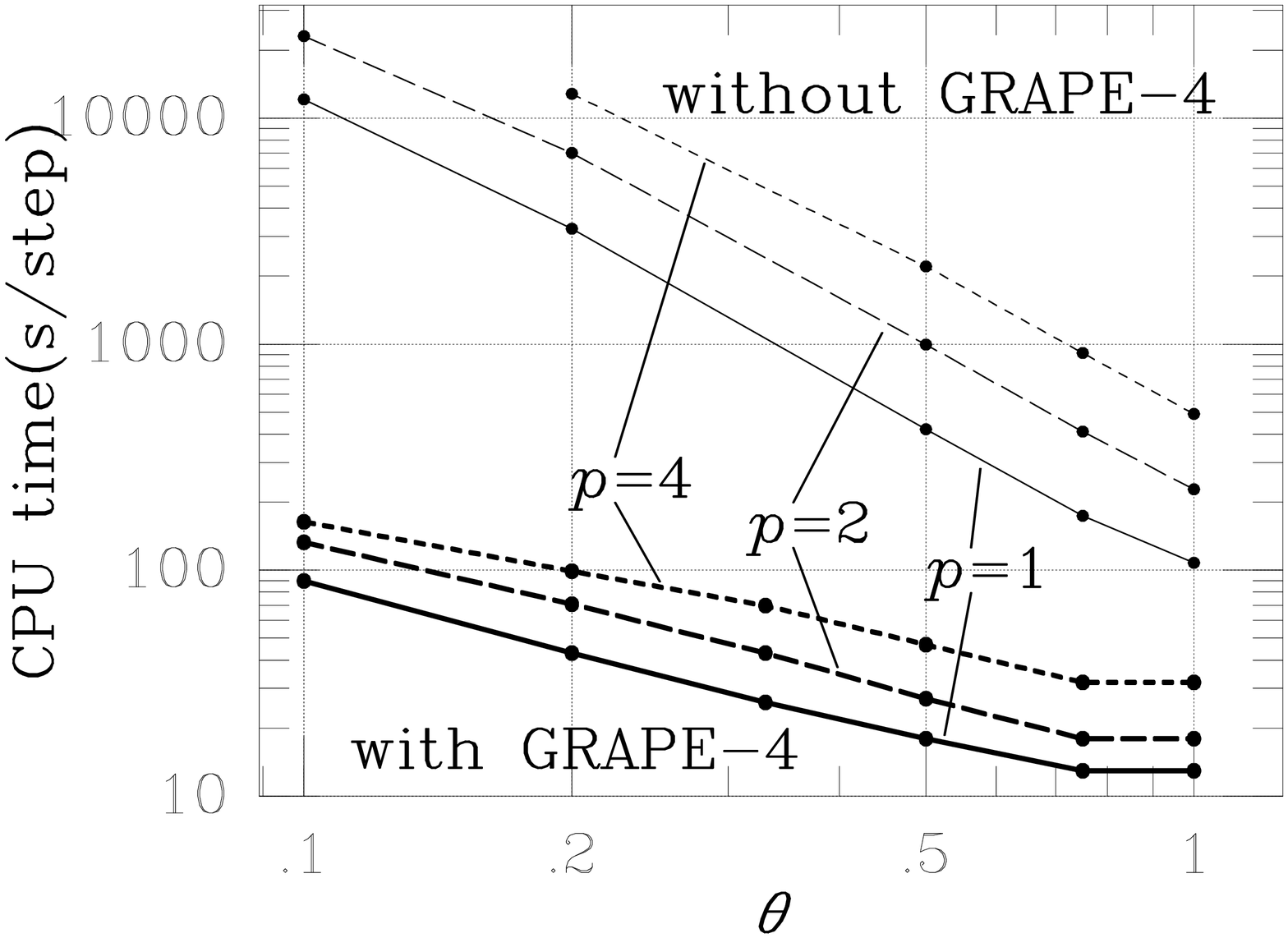}{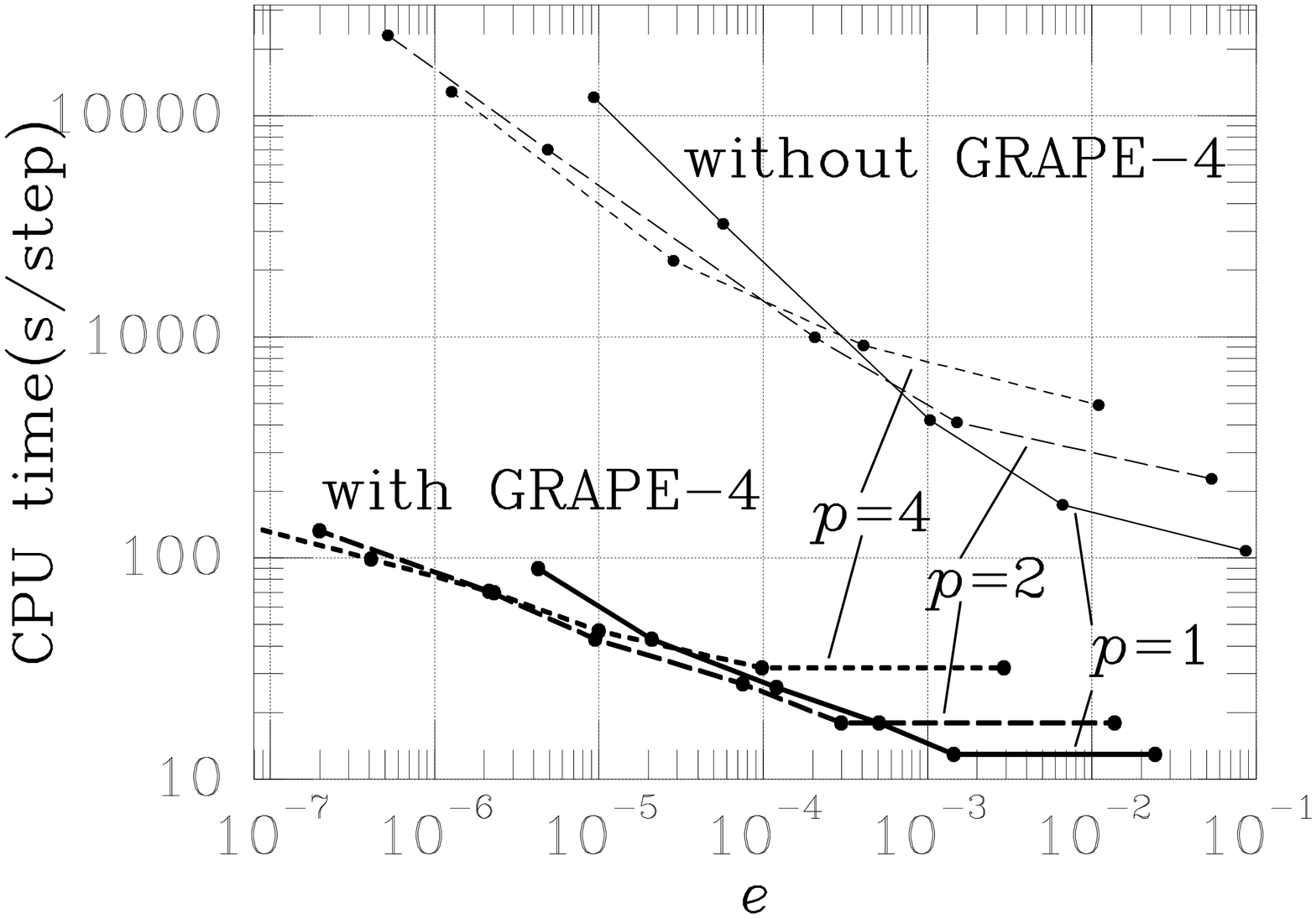}
\caption{Performance of \ppmm ~treecode with and without GRAPE-4. The
CPU time per one timestep is plotted for the opening angle
$\theta$(left panel) and the force calculation error $e$(right panel). 
Thick and thin curves are the results with and without GRAPE-4,
respectively. Solid, long dashed, and short dashed curves are for
$p=1$, $2$, and $4$, respectively.
}
\label{fig:g4}
\end{figure}

Figure \ref{fig:g4} shows the timing results of our treecode.  The CPU
time per timestep is measured on DEC AlphaServer 8400 5/300 (DEC
21164, 300MHz, one processor) with and without GRAPE-4(\cite{m97}). We
used a uniform distribution of 262144 particles in a sphere. The
calculation with GRAPE-4 is 10--100 times faster than that without
GRAPE-4. The gain becomes larger as we increase $p$ or decrease
$\theta$. Thus we can conclude that the combination of \ppmm ~and
GRAPE is quite attractive if high accuracy is required.

\section{Relation between \ppmm ~and Anderson's Method}\label{sec:ppmmam}

\ppmm ~and Anderson's method are quite similar. Both approximate the
multipole expansion by discrete values on points on a sphere. The
difference is that \ppmm ~assigns mass to points while Anderson's
method assigns potential.

\ppmm ~has two advantages over Anderson's method.
One is that the calculation of M2L part is significantly simpler for
\ppmm ~since the expanded potential $\Phi$ is evaluated as the
summation of the potential due to pseudo-particles. The expression of
$\Phi$ in \ppmm ~is given by
\begin{equation}
  \Phi(\Vec{r}) = \sum_{j=1}^K \frac{m_j}{r},
\end{equation}
while the expression for Anderson's method is given by
equation(\ref{eq:amo}). Since M2L is dominant part in FMM calculation,
total calculation speed for the same expansion order would be faster
for \ppmm. As we have seen in section \ref{sec:grape}, \ppmm ~has an
additional advantage that it can make use of special-purpose
computers to achieve further speed up.

The other advantage is that the accuracy of the force calculated with
\ppmm ~is higher than that with Anderson's method for the same
expansion order. In Anderson's method, the potential due to particles
are converted to the potential on the sphere surface. The potential
evaluated during this conversion is expressed as $m_i/r$ and not
truncated to the $p$-th order. Therefore ``aliasing error'' is
introduced into the potential during this conversion. This error tends
to degrade the accuracy. In \ppmm, this aliasing error is completely
suppressed. Of course, it is easy to modify Anderson's method to use
the truncated form. Such a modification would significantly improve
the overall accuracy.

\section{Summary}\label{sec:summary}

In this paper, we present \ppmm, a new method to express multipole
expansion used in FMM and treecode. In \ppmm, multipole expansion is
expressed by pseudo-particles on a sphere.
\ppmm ~has two advantages over the original multipole expansion. One
is its simplicity and the other is that it can effectively use
special-purpose computers. We implemented treecode with \ppmm ~and
evaluated its performance with and without GRAPE-4. We found that
GRAPE-4 accelerates the calculation by a factor of 10-100.
We conclude that \ppmm ~is a quite attractive method to implement FMM
and treecode.

\end{document}